\documentclass[12pt]{article}
\usepackage{geometry,amsmath,amssymb,enumerate,epsfig}
\newcommand{\Iota}{\mathrm{I}}
\newcommand{\emdash}{---}
\newcommand{\mathe}{\mathrm{e}}
\newcommand{\tmem}[1]{{\em #1\/}}
\newcommand{\tmop}[1]{\ensuremath{\operatorname{#1}}}
\newcommand{\tmtextbf}[1]{{\bfseries{#1}}}
\newenvironment{enumeratealpha}{\begin{enumerate}[a{\textup{)}}] }{\end{enumerate}}
\newenvironment{enumerateroman}{\begin{enumerate}[i.] }{\end{enumerate}}
\newenvironment{enumerateromancap}{\begin{enumerate}[I.] }{\end{enumerate}}

\begin{document}

\title{Simulating the All-Order Strong
Coupling Expansion IV:
CP($N - 1$) as a loop model}
\author{Ulli Wolff\thanks{
e-mail: uwolff@physik.hu-berlin.de} \\
Institut f\"ur Physik, Humboldt Universit\"at\\ 
Newtonstr. 15 \\ 
12489 Berlin, Germany
}
\date{}
\maketitle

\begin{abstract}
  We exactly reformulate the lattice CP($N - 1$) spin model on a $D$
  dimensional torus as a loop model whose configurations correspond to the
  complete set of strong coupling graphs of the original system. A Monte Carlo
  algorithm is described and tested that samples the loop model with its
  configurations stored and manipulated as a linked list. Complete absence of
  critical slowing down and correspondingly small errors are found at $D = 2$
  for several observables including the mass gap. Using two different standard
  lattice actions universality is demonstrated in a finite size scaling study.
  The topological charge is identified in the loop model but not yet
  investigated numerically.
\end{abstract}
\begin{flushright} HU-EP-10/01 \end{flushright}
\begin{flushright} SFB/CCP-10-04 \end{flushright}
\thispagestyle{empty}
\newpage

\section{Introduction}

For a few simple lattice field theories Monte Carlo simulation algorithms are
known that are virtually free of critical slowing down. In these cases we have
an almost qualitatively improved control over approaching the universal
continuum or scaling limit which is essential for particle as well as
statistical physics applications. One important line of such developments
started from the Swendsen Wang cluster algorithm {\cite{Swendsen:1987ce}} for
Potts, and in particular Ising models, in arbitrary dimension. In
{\cite{Wolff:1988uh}} the present author has shown how via embedded Ising
spins these techniques can be extended to O($N$) invariant sigma models. This
has led to numerous high precision simulations close to criticality in the
literature. Unfortunately efforts to widen the applicability to other theories
have since ended in frustration in many cases. A well known example concerning
the CP($N - 1)$ model family is discussed in {\cite{Jansen:1991ir}}. Some
theoretical understanding of the restriction to O($N$) is given in
{\cite{Caracciolo:1992nh}}. Significant progress in the simulation of CP($N -
1)$ models has nevertheless been made in the sequel, for example with the
non-recursive multigrid (`unigrid') method in {\cite{Hasenbusch:1992tx}},
{\cite{Hasenbusch:1993na}}. In {\cite{Beard:2004jr}}, {\cite{Beard:2006mq}} an
interesting but somewhat indirect method via quantum models and their
reduction from higher dimension was presented. Nonetheless CP($N - 1)$ systems
seem to remain a challenging testing ground for hopefully more systematically
generalizable attempts to boost the efficiency of the numerical evaluation of
lattice field theory.

A completely different recent research programme is based on the proposal of a
Monte Carlo summation of the strong coupling series (in a certain simple form)
which replaces the sampling of lattice field configurations. For many models
of interest this series converges for any given {\tmem{finite}} volume and
choice of parameters, and the in general infinite set of strong coupling
graphs may be considered as an equivalent non-perturbative representation of
the original lattice model. `Worm' algorithms generalizing those described in
{\cite{prokofev2001wacci}} were the essential tool to allow for an efficient
simulation of this ensemble. In this formulation criticality means that large
graphs are important, which are far too numerous for systematic evaluation.
The question if a Monte Carlo procedure can efficiently sample a sufficient
subset seems rather different from the problem of collectively updating long
distance correlated fields. Little other means than numerical experiments are
presently available to answer this distinct and open question. We here extend
the recent series of papers {\cite{Wolff:2009kp}}, {\cite{Wolff:2008km}},
{\cite{Wolff:2009ke}} to give an affirmative answer also for $\tmop{CP} ( N -
1)$.

In particular in {\cite{Wolff:2009kp}} many details for the treatment of
non-Abelian models have been developed that are similar here. Therefore we
have to constantly refer to this work and the present paper could not really
become self-contained without excessive repetition. On the other hand here
further progress is made for the simulation method that can be used to also
render the O($N$) model simulations even more efficient that what was
described in {\cite{Wolff:2009kp}}.

The present paper is organized as follows. In section 2. we formulate the
CP($N - 1)$ model in several discretizations and derive the equivalent loop
models. For their simulation an algorithm is described in section 3 and
numerically applied in section 4. In section 5. the loop formulation is
generalized to include the $\theta$ parameter coupled to the topological
charge, but related numerical experiments are left to possible future
publications. We end in section 6. with some brief conclusions.

\section{CP($N - 1)$ model as a loop model}

\subsection{Explicit gauge field formulation}

The CP($N - 1$) model is formulated with spins on lattice sites with values in
a complex $N - 1$ dimensional projective space. They may be represented by
one-dimensional projectors or by complex $N$-component unit vectors $\phi (x)
\in \mathbb{C}^N, | \phi | = 1,$ whose phases are irrelevant. One of the
standard lattice actions {\cite{DiVecchia:1981eh}} in use can be written as
\begin{equation}
  - S [\phi, U] = \beta \sum_{x \mu} [U (x, \mu) \phi^{\dag} (x) \phi (x +
  \hat{\mu}) + U^{- 1} (x, \mu) \phi^{\dag} (x + \hat{\mu}) \phi (x)] .
  \label{SU}
\end{equation}
We here sum over all links of a hypercubic periodic lattice in $D$ dimensions
with extent $L_{\mu}$ in direction $\mu = 0, 1, \ldots, D - 1$. We use lattice
units in which the $L_{\mu}$ are integer and $V = \prod L_{\mu}$ is the number
of lattice cells or sites. In addition to the spin field $\phi$ a U(1) gauge
field $U (x, \mu)$ is included which can absorb local phase transformations of
$\phi (x)$. There also is a global SU($N$) invariance with $\phi$ in the
fundamental representation. In a first step we consider $U$ as a given
`background' field over which we integrate later. We define the following
partition function with two contracted adjoint composite insertions
\begin{equation}
  Y [u, v ; U] = \int \left[ \prod_z d \mu (\phi (z)) \right] \mathe^{- S
  [\phi, U]} \phi^{\dag} (u) \lambda^a \phi (u) \phi^{\dag} (v) \lambda^a \phi
  (v) \label{Yuv}
\end{equation}
where $d \mu$ is the normalized invariant measure on the $2 N - 1$ dimensional
sphere of $2 N$ real component unit vectors made from the real and imaginary
part of $\phi$. The generalized Pauli-Gell-Mann matrices $\lambda^a$ with $a =
1, 2, \ldots, N^2 - 1$ are a basis for traceless $N \times N$ matrices that
obey the normalization condition
\begin{equation}
  \tmop{tr} (\lambda^a \lambda^b) = 2 \delta^{a b} . \label{Gell}
\end{equation}
Using the easily proven completeness relation
\begin{equation}
  \lambda^a_{\alpha \beta} \lambda^a_{\gamma \delta} = 2 \left( \delta_{\alpha
  \delta} \delta_{\beta \gamma} - \frac{1}{N} \delta_{\alpha \beta}
  \delta_{\gamma \delta} \right)
\end{equation}
we may verify that our normalization is such that for coinciding $u = v$
\begin{equation}
  Y [u, u ; U] = 2 (1 - 1 / N) \int \left[ \prod_z d \mu (\phi (z)) \right]
  \mathe^{- S [\phi, U]}
\end{equation}
holds and thus the ordinary partition function emerges up to a known factor.

The single site generating function immediately follows from the one of the
O($N$) model {\cite{Wolff:2009kp}}
\begin{equation}
  \int d \mu (\phi) \mathe^{j^{\dag} \phi + \phi^{\dag} j} = \sum_{n =
  0}^{\infty} F [n ; N] (j^{\dag} j)^n \label{Ssite}
\end{equation}
with
\begin{equation}
  F [n ; N] = \frac{\Gamma (N)}{n! \Gamma (N + n)} .
\end{equation}
On each link $x \mu$ ($y = x + \hat{\mu})$ we double-expand
\begin{equation}
  \mathe^{\beta [U \phi^{\dag} (x) \phi (y) + U^{- 1} \phi^{\dag} (y) \phi
  (x)]} = \sum_{k, \overline{k} = 0}^{\infty} \frac{\beta^{k +
  \overline{k}}}{k! \overline{k} !} U^{k - \overline{k}} (\phi^{\dag} (x) \phi
  (y))^k (\phi^{\dag} (y) \phi (x))^{\overline{k}},
\end{equation}
then promote the integers to link fields $k (x, \mu), \overline{k} (x, \mu)$
and also introduce the differences
\begin{equation}
  j_{\mu} (x) \equiv j (x, \mu) = k (x, \mu) - \overline{k} (x, \mu) .
\end{equation}
Auxiliary fields
\begin{equation}
  d (z) = \sum_{\mu} \overline{k} (z - \hat{\mu}, \mu) + k (z, \mu) +
  \delta_{z, u} + \delta_{z, v}
\end{equation}
and
\begin{equation}
  \overline{d} (z) = \sum_{\mu} k (z - \hat{\mu}, \mu) + \overline{k} (z, \mu)
  + \delta_{z, u} + \delta_{z, v}
\end{equation}
count the number of factors $\phi^{\dag}$ and $\phi$ at each site in an
expansion term. The measure (\ref{Ssite}) is U(1) invariant such that nonzero
contributions only result if $d (z) = \overline{d} (z)$ holds at all sites.
This condition is equivalent to the vanishing divergence or flux conservation
\begin{equation}
  d (x) - \overline{d} (x) = \partial_{\mu}^{\ast} j_{\mu} (x) = 0
  \label{divj}
\end{equation}
where $\partial_{\mu}^{\ast}$ is the nearest neighbor backward derivative.

The relevant integral over $\phi$ now is
\begin{equation}
  Y [u, v ; U] = \sum_{k, \overline{k}} \prod_{x \mu} \left[ \frac{\beta^{k
  (x, \mu) + \overline{k} (x, \mu)}}{k (x, \mu) ! \overline{k} (x, \mu) !}
  U^{j (x, \mu)} \right] \int \left[ \prod_z d \mu (\phi (z)) \right] \cdots
\end{equation}
\[ \begin{array}{ll}
     \cdots \phi^{\dag} (u) \lambda^a \phi (u) \phi^{\dag} (v) \lambda^a \phi
     (v) & \prod_{y \nu} (\phi^{\dag} (y) \phi (y + \hat{\nu}))^{k (y, \nu)}
     (\phi^{\dag} (y + \hat{\nu}) \phi (y))^{\overline{k} (y, \nu)} .
   \end{array} \]
Precisely as in {\cite{Wolff:2009kp}} this spin integral can now be performed
by replacing all $\phi, \phi^{\dag}$ by derivatives with respect to sources
$\partial / \partial j^{\dag}$, $\partial / \partial j$ and then using
(\ref{Ssite}) on all sites. All SU($N$) indices get contracted and the various
terms correspond to strong coupling graphs on the lattice. On each link we
draw $k (x, \mu)$ lines with arrows in the positive direction and
$\overline{k} (x, \mu)$ lines with opposite arrows. At each site $z$ we again
imagine all possible ways to saturate all derivatives as a `switchboard' that
connects (contracts) all surrounding lines pairwise with each other, always an
incoming to an outgoing one. This involves the weight $F (d (z) ; N)$ from the
measure at that site. The `connectors' are regarded as 2-vertices joining
pairs of lines at sites. In this way oriented closed loops arise around which
the contractions in internal space contribute a factor $N$ per loop to the
total weight. Two (chains of) lines connect pairwise the four inserted spins
at $u$ and $v$, they end in four 1-vertices. These lines are open
geometrically (unless $u = v$) and are saturated in internal space by the
$\lambda$ matrices. Because these are traceless there is only one nonzero
pairing of the four spins: a positively oriented line pointing from a $u$-spin
[$\phi^{\dag} (u)$] to a $v$-spin [$\phi (v)$] and the other one from $v$ to
$u$. The result of these contractions is a factor $2 (N^2 - 1)$. In analogy to
the O($N$) model in {\cite{Wolff:2009kp}} we call these lines active loops,
distinguished as the $u v$-loop and the $v u$-loop, as opposed to the
remaining passive loops. \ Either active loop is called trivial if for $u = v$
it entirely lives on that site, i.e. it has no lines and 2-vertices. We now
regard $Y [u, v ; U]$ as the sum over all possible loop graphs $\Lambda \in
\overline{\mathcal{L}}_2$ of which each consists of many arbitrarily
overlapping, intersecting and backtracking oriented closed loops plus the two
active loops ending in 1-vertices at $u$ and $v$. At this stage we consider
$k$, $\overline{k}, u, v$ as functions of $\Lambda$. All graphs in
$\overline{\mathcal{L}}_2$ satisfy (\ref{divj}). After counting the
multiplicity of the terms corresponding to each graph in the way discussed in
{\cite{Wolff:2009kp}} we arrive at the remarkably simple form
\begin{equation}
  \mathcal{Y}[U] = C \sum_{u, v} \rho^{- 1} (u - v) Y [u, v ; U] =
  \sum_{\Lambda \in \overline{\mathcal{L}}_2} \rho^{- 1} (u - v) W [\Lambda]
  N^{| \Lambda |} \prod_{x \mu} U^{j (x, \mu)} \label{WU}
\end{equation}
with the weight
\begin{equation}
  W [\Lambda] = \frac1{\mathcal{S}[\Lambda]} \left[ \prod_{x \mu} \beta^{k (x, \mu) + \overline{k} (x,
  \mu)} \right] \prod_z \frac{\Gamma (N)}{\Gamma (N + d (z))}, \hspace{1em}
  C^{- 1} = 2 (1 - 1 / N^2) . \label{Wloop}
\end{equation}
In the exponent $| \Lambda |$ is the number of closed loops in the
configuration $\Lambda$ (including the two active ones in our convention). 
The factor $\mathcal{S}[\Lambda]$ is the symmetry factor of the graph introduced
in the {\em erratum} to {\cite{Wolff:2009kp}}.
The
strictly positive weight $\rho$ with the normalization $\rho (0) = 1$ has been
first discussed in {\cite{Wolff:2008km}} and will be chosen to our convenience
later.

In the standard CP($N - 1$) model one now integrates over all $U (x, \mu)$
independently link by link with the normalized U(1) measure. This enforces the
constraint $j (x, \mu) = 0$ in (\ref{WU}) on all links. We call this subset of
graphs $\mathcal{L}_2^{} \subset \overline{\mathcal{L}}_2^{}$. The flux
represented by the arrows now has to vanish identically on all links instead
of just being conserved at the sites. Then the loop partition function becomes
\begin{equation}
  \mathcal{Z}= \int D U\mathcal{Y}[U] = \sum_{\Lambda \in \mathcal{L}_2^{}}
  \rho^{- 1} (u - v) W [\Lambda] N^{| \Lambda |} . \label{Zloop}
\end{equation}

\subsection{Quartic action formulation}

A second popular action {\cite{DiVecchia:1981eh}} for the CP($N - 1$) model
contains only the field $\phi$ with the action
\begin{equation}
  - S_q [\phi] = 2 \beta_q \sum_{x \mu} | \phi^{\dag} (x) \phi (x + \hat{\mu})
  |^2 . \label{Sq}
\end{equation}
If we start from $S_q$, the same steps as above immediately lead to the
locally \ flux-less ($k \equiv \overline{k}$) graphs $\mathcal{L}_2^{}$
\begin{equation}
  \mathcal{Z}_q = \sum_{\Lambda \in \mathcal{L}_2^{}} \rho^{- 1} (u - v) W_q
  [\Lambda] N^{| \Lambda |} \label{Zloopq}
\end{equation}
with the weight, modified by the slightly different multiplicities,
\begin{equation}
  W_q [\Lambda] = \frac1{\mathcal{S}[\Lambda]} \left[ \prod_{x \mu} [2 \beta_q]^{k (x, \mu)} k (x, \mu) !
  \right] \prod_z \frac{\Gamma (N)}{\Gamma (N + d (z))} . \label{Wq}
\end{equation}

For either action the relation between the two point correlation of the
original spin model and the ensemble (\ref{Zloop}) or (\ref{Zloopq}) is easy
to establish. With double angle expectation values defined by [$W \rightarrow
W_q$ for (\ref{Zloopq})]
\begin{equation}
  \langle \langle \mathcal{O}(\Lambda) \rangle \rangle = \frac{1}{\mathcal{Z}}
  \sum_{\Lambda \in \mathcal{L}_2^{}} \rho^{- 1} (u - v) W [\Lambda] N^{|
  \Lambda |} \mathcal{O}(\Lambda)
\end{equation}
we find
\begin{equation}
  \langle \phi^{\dag} (0) \lambda^a \phi (0) \phi^{\dag} (x) \lambda^b \phi
  (x) \rangle = \rho (x) \frac{2 \delta_{a b}}{N (N + 1)}  \frac{\langle
  \langle \delta_{u - v, x} \rangle \rangle}{\langle \langle \delta_{u, v}
  \rangle \rangle} . \label{corr2}
\end{equation}
In particular the susceptibility
\begin{equation}
  \chi = \frac{1}{2} \sum_x \langle \phi^{\dag} (0) \lambda^a \phi (0)
  \phi^{\dag} (x) \lambda^a \phi (x) \rangle = \sum_x \left\{ \langle
  \tmop{tr} [\phi (0) \phi^{\dag} (0) \phi (x) \phi^{\dag} (x)] \rangle -
  \frac{1}{N} \right\}
\end{equation}
can be measured as
\begin{equation}
  \chi = \frac{N - 1}{N} \frac{\langle \langle \rho (u - v \rangle
  \rangle}{\langle \langle \delta_{u, v} \rangle \rangle} . \label{chiobs}
\end{equation}

\subsection{Adjoint formulation}

The action $S_q$ may also be rewritten completely in terms of adjoint
composite fields
\begin{equation}
  J^a (x) = \phi^{\dag} (x) \lambda^a \phi (x) .
\end{equation}
We introduce
\begin{equation}
  - S_a [\phi] = \beta_a \sum_{x \mu} J^c (x) J^c (x + \hat{\mu})
\end{equation}
but note that in the path integral we integrate over $\phi (x)$ as before.
There is the trivial relation, again a consequence of (\ref{Gell}),
\begin{equation}
  S_a [\phi] = S_q [\phi] |_{\beta_q = \beta_a} + \frac{2 D V}{N} \beta_a .
\end{equation}
Thus $S_q$ and $S_a$ just differ by a constant shift, irrelevant for any
correlation. In spite of this the all-order strong coupling expansions are far
from identical. An expansion in powers of $\beta_a$ using $S_a$ would involve
unoriented lines as for the O($N$) model. A major difference arises however
from the measure that would be relevant in this case
\begin{equation}
  \int d \mu (\phi) \mathe^{b^a J^a} = \sum_{m, n = 0}^{\infty} A [m, n ; N]
  (b^a b^a)^m (d_{a b c} b^a b^b b^c)^n \label{Ssitead}
\end{equation}
with a source $b^a$. In the adjoint representation there is (for $N > 2$) a
second invariant totally symmetric tensor beside $\delta_{a b}$, namely
\begin{equation}
  d_{a b c} = \frac{1}{4} \tmop{tr} (\lambda^a \lambda^b \lambda^c + \lambda^b
  \lambda^a \lambda^c) .
\end{equation}
Thus this expansion is more complicated with both two point and three point
vertices available to contribute. Only for $N = 2$ $d_{a b c}$ vanishes and
the adjoint formulation falls back to the standard lattice formulation of the
O($3 = N^2 - 1)$ model as is well known. But even here $S_q$ yields a new
expansion and the numerical reproduction of O(3) results in the CP(1)
formulation is non-trivial.

\subsection{Nienhuis Boltzmann factor}

In {\cite{Wolff:2009kp}} we have locally modified the original Boltzmann
factor by truncating its expansion on each bond by $k (x, \mu) \leqslant
k_{\max}$. The original model is recovered for large $k_{\max}$ but a
particularly interesting case arises for $k_{\max} = 1$. Such modifications
and the conjecture of universality still holding have been introduced into the
literature a long time ago {\cite{Domany:1981fg}}, {\cite{Nienhuis:1982fx}}.
Truncating in the $S_q$ formulation amounts to the replacement
\begin{equation}
  \mathe^{- S_q [\phi]} \rightarrow \mathcal{N}_q [\phi] = \prod_{x \mu}
  \left[ 1 + 2 \tilde{\beta}_q | \phi^{\dag} (x) \phi (x + \hat{\mu}) |^2
  \right] . \label{Niebol}
\end{equation}
The adjoint translation reads
\begin{equation}
  \mathcal{N}_q [\phi] = (1 + 2 \tilde{\beta}_q / N)^{D V} \prod_{x \mu}
  \left[ 1 + \tilde{\beta}_a J^a (x) J^a (x + \hat{\mu}) \right]
\end{equation}
with the adjoint Nienhuis coupling
\begin{equation}
  \tilde{\beta}_a = \frac{\tilde{\beta}_q}{1 + 2 \tilde{\beta}_q / N} .
\end{equation}
For $\tilde{\beta}_q \geqslant 0$ we find $0 \leqslant \tilde{\beta}_a
\leqslant N / 2$. The criticality observed for O(3) in {\cite{Wolff:2009kp}}
corresponds to large positive $\tilde{\beta}_a$ and $N = 2$. This is obviously
not reached for positive $\tilde{\beta}_q$ but only as $\tilde{\beta}_q
\nearrow (- N / 2)$. We now turn to the possibilities of efficient Monte Carlo
algorithms for the CP($N - 1)$ loop model just outlined. We will however find
this to be restricted to positive $\beta_q$ and hence cannot at present
implement the Nienhuis formulation as for the O(3) model.

\section{Simulation of the CP($N - 1)$ loop model}

\subsection{Algorithm R for real $N$}

We have arrived at a representation of the CP($N - 1)$ model in terms of loops
representing arbitrary strong coupling graphs. These configurations can be
parameterized by linked lists as described in detail in {\cite{Wolff:2009kp}}
with only minor adjustments. One of the two possible orientations of each loop
(for example column 2 of the list) is identified now with the physical
orientation of lines of the present graphs. The flag in column 4 must now
allow for three different values to distinguish between passive loops, the $u
v$ loop and the $v u$ loop.

We first develop an algorithm to simulate the CP($N - 1$) loop ensemble
(\ref{Zloop}). We define a number of separate update steps such that each of
them fulfills detailed balance. They will then be iterated in some order as
the final update procedure. The moves are all Metropolis proposals for which
we quote a ratio $q$ which yields the acceptance probability $\min (1, q)$. We
encounter cases where our a priori proposal probabilities are not symmetric.
To achieve detailed balance this needs to be compensated in $q$, which is then
not just the ratio of the Boltzmann weights of the two configurations
involved. This in particular applies to the inclusion of $\mathcal{S}[\Lambda]$
as discussed in the erratum to {\cite{Wolff:2009kp}}.
\begin{enumerateromancap}
  \item Extension and retraction step: With probability $p_{\tmop{ext}} = 1 /
  (1 + r_{\tmop{ext}})$ we propose an extension step, otherwise the retraction
  step. In the extension branch we next choose with equal probability one of
  the $2 D$ directions to move $u$ to the corresponding neighbor $\tilde{u}$
  with a concurrent extension of both active loops. The proposal is accepted
  according to the ratio
  \begin{equation}
    q_{\tmop{ext}} = \frac{2 D \beta^2 r_{\tmop{ext}}}{(N + d (
    \text{$\tilde{u}$)}) (N + d ( \text{$\tilde{u}$}) + 1)}  \frac{\rho (u -
    v)}{\rho ( \tilde{u} - v)} . \label{qext}
  \end{equation}
  In the retraction branch $u$ is pulled back over one link along the active
  loops with the ratio
  \begin{equation}
    q_{\tmop{ret}} = \frac{(N + d ( \text{$u$}) - 1) (N + d ( \text{$u$}) -
    2)}{2 D \beta^2 r_{\tmop{ext}}}  \frac{\rho (u - v)}{\rho ( \tilde{u} -
    v)} \label{qret}
  \end{equation}
  if {\tmem{both active loops}} lead to the same neighbor $\tilde{u}$,
  otherwise no move is made.
  
  \item Re-route step: We here want to change the SU($N$) contraction or line
  connectivity structure at $u$. Such a move is only considered here if $u
  \not= v$ and $d (u) > 2$ holds.
  \begin{enumerateroman}
    \item We pick one of the 2-vertices at $u$ and propose to swap between the
    line pointing out of this 2-vertex and the $u v$ loop emerging from its
    initial 1-vertex at $u$. For the $q$ ratio we need to distinguish further
    sub-cases.
    \begin{enumeratealpha}
      \item The chosen 2-vertex belongs to a passive loop. The latter then
      effectively gets inserted into the $u v$ active loop at $u$, $| \Lambda
      |$ is reduced by one and we accept/reject with $q$=$1 / N$.
      \item The chosen 2-vertex belongs to the $u v$ loop itself, which
      self-intersects at $u$. Then a section is detached from it forming a new
      passive loop, $| \Lambda |$ goes up by one, $q = N$.
    \end{enumeratealpha}
    No move is made if the chosen 2-vertex belongs to the $v u$ loop.
    \item As i. but the r\^ole of of the two active loops switched.
  \end{enumerateroman}
\end{enumerateromancap}
\begin{figure}[htb]
\begin{center}
  \resizebox{0.4\textwidth}{!}{\includegraphics{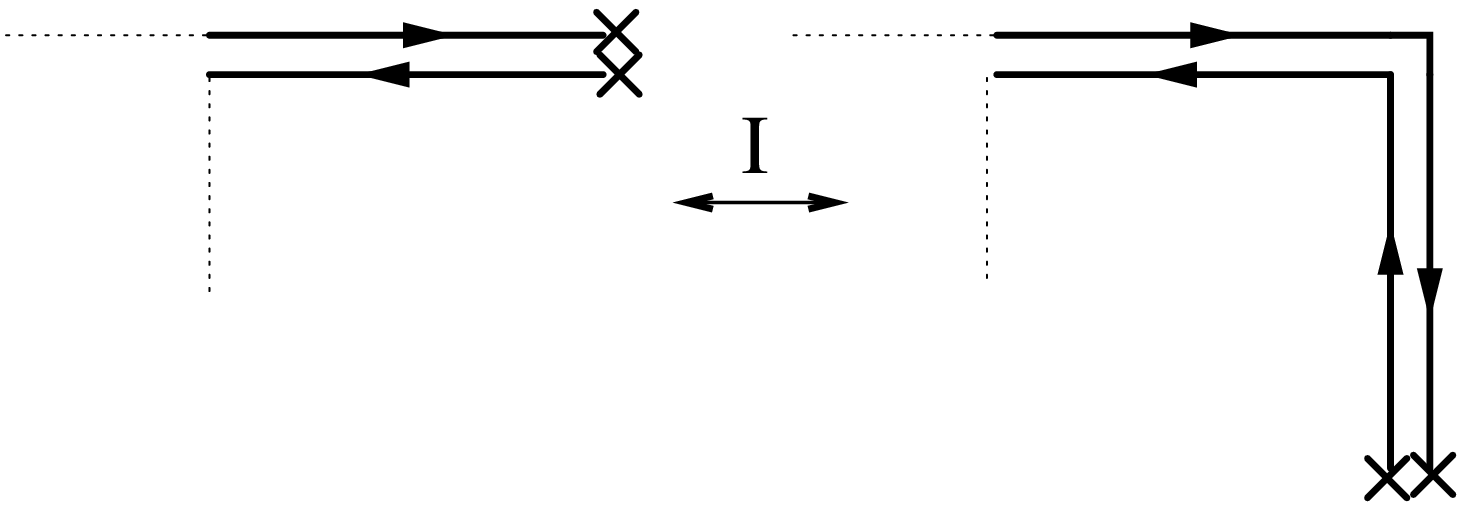}}
  \quad
  \resizebox{0.55\textwidth}{!}{\includegraphics{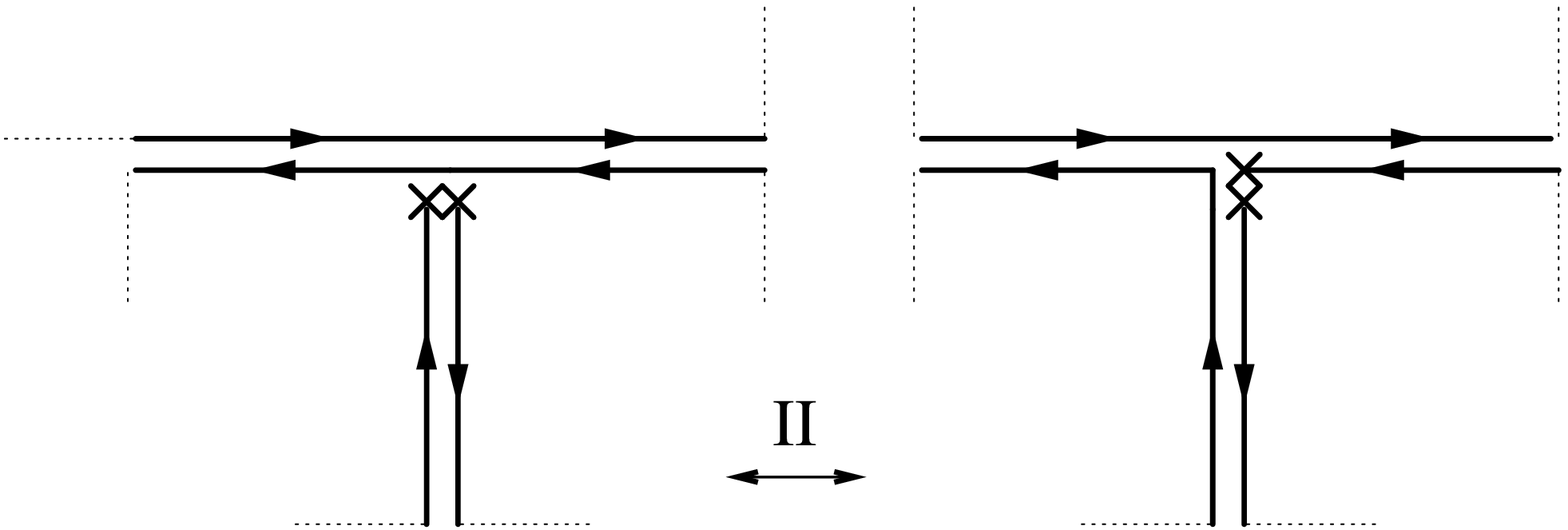}}
  \caption{Illustration of elementary update steps I (left) and II (right).
  Dotted lines indicate the continuation with other parts of a graph
  $\Lambda$. The crosses are at the insertion site $u$.
  \label{steps}}
\end{center}
\end{figure}
In figure \ref{steps} we try to graphically illustrate the elementary moves.
We iterate these steps {\emdash} and similar ones focussing on $v$ instead of
$u$ {\emdash} according to the scheme
\begin{equation}
  1 \tmop{Iteration} = (\Iota_u \Iota \Iota_u \Iota_v \Iota \Iota_v)^{N \times
  V / 2} . \label{Rit}
\end{equation}
The ergodicity of this algorithm is shown as usual. One has to convince
oneself that any graph can be built starting from the empty one by repeating
the above moves. The empty graph is indeed the configuration from which we
will start all simulations.

For a simulation with the quartic action (\ref{Sq}) we just have to
replace (\ref{qext}) by
\begin{equation}
  q_{\tmop{ext}}' = \frac{4 D \beta_q r_{\tmop{ext}} (k (l) + 1)}{(N + d (
  \text{$\tilde{u}$)}) (N + d ( \text{$\tilde{u}$}) + 1)}  \frac{\rho (u -
  v)}{\rho ( \tilde{u} - v)}
\end{equation}
and (\ref{qret}) by
\begin{equation}
  q_{\tmop{ret}}' = \frac{(N + d ( \text{$u$}) - 1) (N + d ( \text{$u$}) -
  2)}{4 D \beta_q r_{\tmop{ext}} k (l)}  \frac{\rho (u - v)}{\rho ( \tilde{u}
  - v)} .
\end{equation}
In the last two formulae $l$ denotes the link between $u$ and $\tilde{u}$,
i.e. $l = (u, \mu)$ if $\tilde{u} = u + \hat{\mu}$ and $l = ( \tilde{u}, \mu)$
if $\tilde{u} = u - \hat{\mu}$ with $\hat{\mu}$ denoting a unit vector in the
positive $\mu$ direction.

In comparison with {\cite{Wolff:2009kp}} a few changes can be noticed. The
probability $p_{\tmop{ext}}$ is new here. In {\cite{Wolff:2009kp}}\, we have
effectively chosen the special value $p_{\tmop{ext}} = 1 / (2 D + 1)$. \ We
found the greater flexibility here useful to prevent acceptance rates from
getting small. In all runs reported about below we have taken $p_{\tmop{ext}}
= 0.3$ and found unproblematic high acceptance rates. In addition there are no
analogs of the steps IIi, IIii and III of {\cite{Wolff:2009kp}}. We found that
they are not essential and also the O($N$) algorithm may be simplified
correspondingly. As a more technical change we have slightly modified the
handling of the threefold linked list. As before two linkages allow to travel
along loops in both directions while a third set of pointers allows to
efficiently enumerate the vertices at a given lattice site. As we delete
vertices (free list entries) in the retract step, we now immediately re-adjust
all pointers including the site related ones.

\subsection{Algorithm I for integer $N$}

\subsubsection{Why an improved method?}

We have extensively run the algorithm just described and successfully
reproduced numbers from the literature {\cite{Wolff:1992ze}}
{\cite{Hasenbusch:1992tx}} {\cite{Hasenbusch:1993na}} for small and
intermediate lattices ($L < 200$ in $D = 2$, $N = 4$) with permille errors
using only moderate PC time. As for the O($N$) simulations in
{\cite{Wolff:2009kp}} we have found that integrated autocorrelation times
{\tmem{in units of iterations}} typically stay below one in our timeseries of
$10^6$ iterations.

There nevertheless is a problem that is exacerbated here compared to O($N$).
The moves described above seem to be superficially local with a fixed number
of operations around $u$ and $v$. For the reconnect II we have to know however
if the 2-vertex that has been randomly chosen is part of a passive or an
active loop, which represents nonlocal information that is locally available
to us in the flag entry of the list {\cite{Wolff:2009kp}}. The price to pay
for this is that whenever sections of loops are detached or eaten up, these
have to re-flagged. It turns out that thus, close to the continuum limit on
large lattices, most of the CPU time is spent travelling around loops
resetting flags. This problem was already discussed in {\cite{Wolff:2009kp}}.
It is more severe for CP($N - 1$) than for O($N$) because the loops are more
numerous and probably longer for comparable correlation lengths and lattice
sizes.

To diagnose the problem in more detail we have measured the loop-size
distribution
\begin{equation}
  \ell_n = \Big\langle \Big\langle \sum_{i = 1}^{| \Lambda |} \delta_{| \lambda_i |,
  n} \Big\rangle \Big\rangle_0
\end{equation}
where the expectation value
\begin{equation}
  \langle \langle \mathcal{O}(\Lambda) \rangle \rangle_0 = \frac{\langle
  \langle \mathcal{O}(\Lambda) \delta_{u, v} \rangle \rangle}{\langle \langle
  \delta_{u, v} \rangle \rangle}
\end{equation}
refers to the closed `vacuum' graphs only. Further we have introduced and
implemented the decomposition of each such graph $\Lambda$ into individual
closed loops
\begin{equation}
  \text{$\Lambda = \bigcup_{i = 1}^{| \Lambda |} \lambda_i$}
\end{equation}
of lengths (number of link-lines) $| \lambda_i |$. In {\cite{Janke:2009rb}},
inspired by percolation and random walk theory, the asymptotic scaling form
\begin{equation}
  \ell_n \propto n^{- \tau} \mathe^{- \theta n}
\end{equation}
has been proposed introducing the loop tension $\theta$ and a power correction
exponent $\tau$. We found that such a fit describes reasonably well data that
we have generated with $S_q$ at $N = 4$ for correlations lengths $m^{- 1} = 2
\ldots .18$ and size $m L \approx 10$. Our aim here was not a high precision
estimation of $\theta$ and $\tau$. A clean assessment of their systematic
errors would require quite some effort as it depends on the chosen fit window
and their mutual correlation. We content ourselves at present with quoting
that our analysis suggests that our data are roughly described by
\begin{equation}
  \tau \approx 1.8, \hspace{1em} \theta \approx m^2 / 17.
\end{equation}
If we now pick a random 2-vertex we hit long loops more frequently and it will
be part of a loop of length $n$ with a probability proportional to $n \ell_n$.
The cost to re-flag such a loop will hence on average be
\begin{equation}
  \frac{\text{$\sum_n n^2 \ell_n$}}{\sum_k k \ell_k} \approx (2 - \tau)
  \theta^{- 1} .
\end{equation}
Because $\theta^{- 1}$ grows roughly proportional to the {\tmem{squared
correlation length}} in the continuum limit (`fractal dimension two' )
simulations with the R algorithm slow down asymptotically in a way similar to
ordinary local methods. It is to be noted that this section of the code (the
re-flag while-loop) is very simple and thus dominates on larger lattices only,
and that loop simulations still have great advantages. In any case we have
found that the CPU time per site for R grew roughly by a factor 60 as the
correlation length and $L$ are scaled up by a factor 7. Slower memory access
on larger lattices may however also have entered here to some degree.

\subsubsection{Elimination of slowing down}

To proceed it was instructive to draw some analogies to Fortuin-Kastelyn
cluster based algorithms for the $q$-state Potts model. In an early proposal
Sweeny {\cite{PhysRevB.27.4445}} has worked with bond variables only. The
Boltzmann weight then contains a factor $q^{N_c}$, where $N_c$ is the number
of percolation clusters implied by the bond configuration. It resembles our
weight $N^{| \Lambda |}$, for instance by allowing to analytically continue in
$q$. During a local bond update nonlocal information is required: does the
status of the single bond change $N_c$ or not? This leads to the same kind of
slowing down. Sweeny has designed a system of hierarchical express pointers to
accelerate the travel around loops (like express trains of the New York
subway). One could consider such an improvement also for R here.

A much simpler solution of the problem was however the one of Swendsen and
Wang {\cite{Swendsen:1987ce}}. They keep spin {\tmem{and}} bond variables and
update them in alternating order. One may actually view the cluster-wise
assigned spins as just a device to stochastically dissolve the nonlocal weight
into local steps. Transferring this technique to our problem at hand we
insert, for example in (\ref{Zloop}), the representation
\begin{equation}
  \text{$N^{| \Lambda |}$} = \sum_{\alpha_1 = 1}^N \ldots \sum_{\alpha_{|
  \Lambda |} = 1}^N 1 = \frac{N}{N - 1} {\sum_{\alpha}}' 1.
\end{equation}
Here the enlarged phase space now consists of graphs $\Lambda$ where each loop
$\lambda_i$ contained in it carries a label $\alpha_i$ that is freely summed
over $N$ values. The update is extended now to such configurations including a
choice for all $\alpha_i$. In the last (primed) sum we omit $\alpha$
assignments where the indices of the two active loops coincide. The reason for
this small extra twist will become clear soon. The loop partition function now
reads for example
\begin{equation}
  \mathcal{Z}= \int D U\mathcal{Y}[U] = \frac{N}{N - 1} \sum_{\Lambda \in
  \mathcal{L}_2^{}} {\sum_{\alpha}}' \rho^{- 1} (u - v) W [\Lambda]
  \label{ZloopI}
\end{equation}
and expectation values in this ensemble are formed in the obvious way.

The simulation of the ensemble including the $\alpha$ values requires only
minimal changes. The flag entries of the linked list are rededicated to now
store the flavor values $\alpha \in \{1, 2, \ldots, N\}$ of each vertex that
is inherited from the loop to which it belongs. The extend/retract step I is
completely unchanged, flavor is just passed on in extensions. The re-route
step II (at fixed $\alpha$) becomes even simpler now and proceeds as follows:
\begin{itemize}
  \item[II'] If $u \not= v$ and $d (u) > 2$ holds we pick one of the 2-vertices
  at $u$. If it carries the same flavor as the initial 1-vertex of the $u v$
  loop we (always) swap in the way described before, otherwise nothing is
  done. Then the same is repeated for $v u$ loop at $u$.
\end{itemize}
Because of the restriction $\alpha_{u v} \neq \alpha_{v u}$ it never happens
that a re-route now leads to a line connecting for example $\phi^{\dag} (u)$
with $\phi (u)$, a contribution not included in our definition of the class
$\mathcal{L}_2$ or $\overline{\mathcal{L}}_2$ designed such that $u, v$ map
out the {\tmem{adjoint}} correlation. The larger class would\, lead to a
correlation with a singlet part decaying to a known constant instead of zero.
It could be canceled but we would presumably be left with more and unnecessary
noise.

For ergodicity we also have to periodically update the flavor assignments
after a certain number of the steps just discussed. We expect the cost for one
iteration to remain O($V$) if we do this only after O($V$) extend/retract
steps. One obvious option for these new steps would be to identify all loops
$\lambda_i$ in $\Lambda$ (including $u v$ and $v u$) and re-flavor them
randomly. Some thought shows however that for ergodicity it is already
sufficient to only randomly re-flavor the two active loops to one of the $N (N
- 1)$ pairs $\alpha_{u v} \neq \alpha_{v u}$. We call this type of re-flavor step
now III. Some short experiments identified the following combination as quite
efficient \
\begin{equation}
  1 \tmop{Iteration} = [(\Iota_u \Iota \Iota_u \Iota_v \Iota \Iota_v)^{V / 2}
  \tmop{III}]^N .
\end{equation}
In particular our experiments have shown that it is {\tmem{not}} profitable to
re-flavor all loops. Most important, it was quite pleasant to find that the
errors of relevant observables grow only by 10\% or so if they are accumulated
during the same number of such I-Iterations replacing the much more costly
R-Iterations. A more detailed discussion of autocorrelations and (the absence
of) critical slowing down will follow in the next section.

\section{Numerical tests}

We here report data on a number of simulations in $D = 2$ dimensions. The
reason for this restriction is that physical interest in the CP($N - 1$)
models and hence available data seem to be concentrated on this
dimensionality. We expect our new formulation and algorithms to be
generalizable to other dimensions without problems.

We have implemented the algorithms R and I as C codes. The graphs $\Lambda$
are encoded into a linked list as described in detail in
{\cite{Wolff:2009kp}}. In C a very natural implementation uses structures of
pointers that are dynamically created and erased. We have found however that
the more static storage handling described in {\cite{Wolff:2009kp}} offers
slight advantages in speed and have hence returned to it in the final version.

\subsection{Validation with the action $S_q$ in large volumes}

We first simulate the action $S_q$ for $N = 4$ to become able to compare (and
find consistency) with data in \ {\cite{Wolff:1992ze}}
{\cite{Hasenbusch:1992tx}} {\cite{Hasenbusch:1993na}} \ on a lattice by
lattice basis. These simulations are summarized in table \ref{tab1}.
\begin{table}[htb]
\begin{center}
  \begin{tabular}{|l|l|l|l|l|l|l|}
    \hline
    $\beta_q$ & $L$ & $\chi$ & $\xi_1$ & $\xi_2$ & $L / \xi_2$ & $K$\\
    \hline
    2.3 & 20 & 11.064(9) & 2.5736(12) & 2.6045(10) & 7.7 & 2.59585(24)\\
    \hline
    2.5 & 32 & 26.214(28) & 4.4491(27) & 4.5097(23) & 7.1 & 3.07843(18)\\
    \hline
    2.7 & 64 & 79.57(11) & 8.7813(76) & 8.9018(61) & 7.2 & 3.57230(10)\\
    \hline
    2.9 & 128 & 275.13(50) & 18.521(22) & 18.837(18) & 6.8 & 4.03968(6)\\
    \hline
    3.1 & 256 & 930.2(2.2) & 38.087(62) & 38.639(50) & 6.6 & 4.48085(3)\\
    \hline
    3.3 & 512 & 2998.6(8.2) & 74.80(16) & 75.54(13) & 6.8 & 4.90816(2)\\
    \hline
  \end{tabular}
  \caption{Results of simulations with the quartic action (\ref{Sq}) that are
  immediately comparable to published data.\label{tab1}}
\end{center}
\end{table}
After each extend/retract we have continuously recorded the
contributions to (\ref{chiobs}) and to the time slice correlations
\begin{equation}
  G (t) = \langle \langle \rho (u - v) [\delta_{t, u_0 - v_0} + \delta_{t, u_1
  - v_1}] \rangle \rangle
\end{equation}
where the $\delta$ functions are $L$-periodic and we enhance the statistics by
summing over both directions for $L_0 = L_1 \equiv L$. As discussed in
{\cite{Wolff:2009kp}} (see also {\cite{Wolff:2008km}}) we took $\rho (x)$
proportional to the free lattice propagator with a mass $\hat{M}$ close to the
one expected for the simulated lattice. From successive pairs of time slices
we determine an effective mass by solving
\begin{equation}
  \frac{G (t + 1)}{G (t)} = \frac{\cosh (m (t + 1 - L / 2)}{\cosh (m (t - L /
  2)}, \hspace{1em} \rightarrow m = m_{\tmop{eff}} (t + 1 / 2) .
\end{equation}
Following {\cite{Wolff:1992ze}} correlation lengths $\xi_k = 1 /
m_{\tmop{eff}} (t_k + 1 / 2)$ are determined self-consis\-tently such that $k
\xi_k \in [t_k, t_k + 1]$, i.e. roughly at separations $\xi$ and $2 \xi$.

In the column
\begin{equation}
  K = \frac{1}{D V} \Big\langle \Big\langle \sum_{x \mu} k (x, \mu) 
                    \Big\rangle \Big\rangle_0
\end{equation}
we list the average number of lines per link in the simulated graphs. Actually
there are $K$ lines of either orientation. As in the O($N$) model (see
{\cite{Wolff:2009kp}}) $K$ is a direct measure of the internal energy due to
the identity
\begin{equation}
  K = 2 \beta_q \langle | \phi^{\dag} (x) \phi (x + \hat{\mu}) |^2 \rangle .
\end{equation}

Each row in table \ref{tab1} derives from $10^7$ iterations of the I
algorithm. We have stored our data as $10^5$ blocks from 100 successive
iterations each. With these blocks an ordinary error analysis with the tool
{\cite{Wolff:2003sm}} was carried out. Between these measurements, on average
separated by 100 iterations, hardly any autocorrelations are detectable. The
errors of our errors are thus at a percent level and all digits given in the
tables are significant. In addition, using multicore PCs, we always simulate
between 8 and 32 independent replica and monitor for acceptable Q-values
{\cite{Wolff:2003sm}}.

To extract the number of loops $| \Lambda |$ in I-simulations one has to
implement an additional observable to obtain this information. In our original
R-simulations it is is however available `for free' and was measured. We have
found values $| \Lambda | / V = 1.25 \ldots .1.05$ slowly falling as $\beta$
rises, thus O(1) loop per site.
\begin{figure}[htb]
\begin{center}
  \resizebox{0.9\textwidth}{!}{\includegraphics{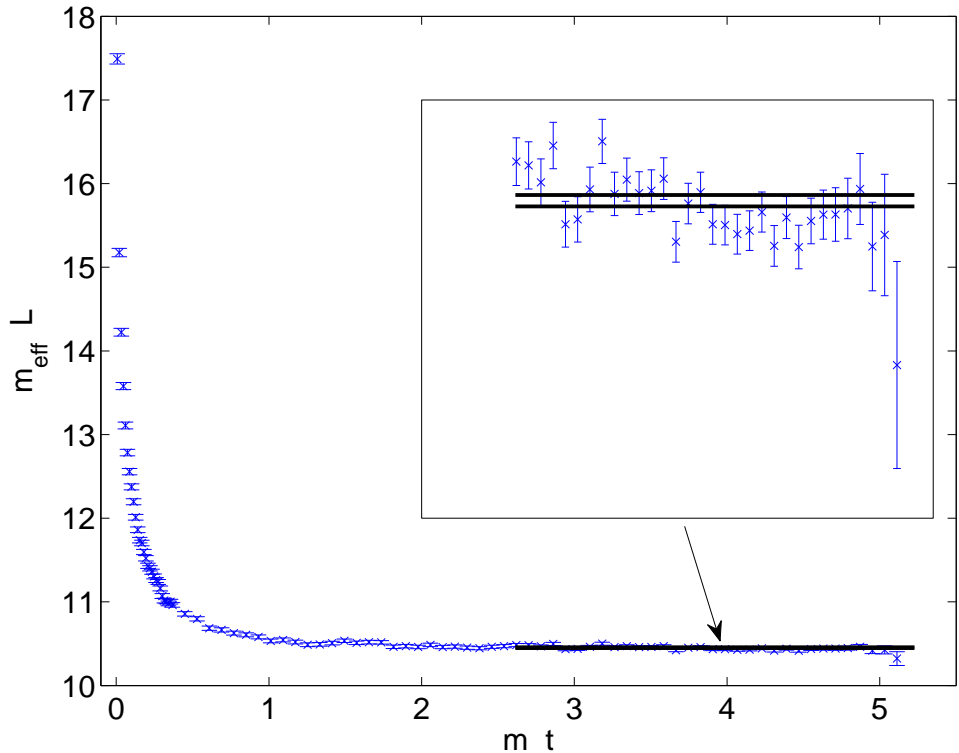}}
  \caption{Effective mass as a function of time slice separation at $\beta_q =
  3.3$ and $L = 780$.\label{fig1}}
\end{center}
\end{figure}
The `historic' parameter sets in table \ref{tab1} have led to physical sizes
of about $m L \approx 7$. We have found that this is too small to see a
convincing mass-plateau at our present precision level. We have therefore
repeated all runs on larger lattices $m L \approx 10$, see table \ref{tab2}.
In figure \ref{fig1} the effective mass on the largest lattice now shows a
satisfactory plateau. Beyond the steep initial decay, we show only every sixth
effective mass value to not clutter up the plot. The size of the relative
statistical errors of $m_{\tmop{eff}}$ is separation independent as expected
{\cite{Wolff:2008km}}. The growth very close to $t = L / 2$ has kinematic
reasons since at $L / 2$ the correlation has a minimum for any mass. We then
decided on these larger lattices to quote mass values from a fit to the
correlation over a window $t = L / 4$ to $ t = L / 2$. This is the horizontal
line errorband in figure \ref{fig1}. We refer to {\cite{Wolff:2009kp}} for a
discussion of the fitting procedure which we took over unchanged.
\begin{table}[htb]
\begin{center}
  \begin{tabular}{|l|l|l|l|l|}
    \hline
    $\beta$ & $L$ & $\chi$ & $\xi (L / 4 \rightarrow L / 2)$ & $K$\\
    \hline
    2.3 & 26 & 11.015(9) & 2.5889(7) & 2.59502(19)\\
    \hline
    2.5 & 44 & 26.015(29) & 4.4561(13) & 3.07771(14)\\
    \hline
    2.7 & 88 & 78.99(12) & 8.8256(28) & 3.57190(8)\\
    \hline
    2.9 & 188 & 272.78(53) & 18.574(7) & 4.03950(4)\\
    \hline
    3.1 & 380 & 926.4(2.4) & 38.068(15) & 4.48080(2)\\
    \hline
    3.3 & 780 & 2960.2(8.6) & 74.619(32) & 4.90818(1)\\
    \hline
  \end{tabular}
  \caption{Data from simulations of the CP(3) model with $m L = 10$. The mass
  $m = 1 / \xi$ is determined by a fit to the correlation. \label{tab2}}
\end{center}
\end{table}
The lattices of table \ref{tab2} represent a scaling set where the
lattice spacing changes toward the continuum \ `at fixed physics'. The weight
$\rho$ was formed with $\hat{M} = 10 / L$. The optimal $\rho$ is observable
dependent. We repeated the run $L = 380$ with $\rho \equiv 1$. The resulting
error in $\chi = 922.89 (57)$ is about four times smaller while the error in
$\xi = 38.089 (33)$ went up by more than a factor two. The execution time
{\tmem{per site}} of an I iteration between the first and the last line of
table \ref{tab2} still goes up by a (modest) factor 1.7, which is probably an
effect of memory or cache access with 8 replica (cores) sharing the memory.
The total CPU time of the run $L = 780$ was about 330 hours on one
double-quadcore Xeon (2.27GHz).

We close on a warning side-remark. With all our observables being positive we
originally thought that a single precision simulation (about 20 \% faster)
would be sufficient. We then found however that in plots like figure
\ref{fig1} the errorbars scattered around a smooth curve to a degree that
seemed implausible. With the transition to double precision throughout (it was
always used for the data analysis) this effect immediately went away. The mean
value of the mass in single precision was still compatible with the new value
while $\chi$ was found to be different beyond errors.

\subsection{Autocorrelations}

In ordinary simulations the errors of observables are influenced by both the
variance of the quantity and by its integrated autocorrelation time. The
latter depends on the algorithm in use while the former is a property only of
the ensemble to be simulated. This distinction tends to get blurred somewhat
for our strong coupling simulations as already discussed in
{\cite{Wolff:2008km}}. The point is that histograms like $\mathcal{O}(x) =
\langle \langle \delta_{x, u - v} \rangle \rangle$ in the simplest case are
measured continuously during the update. An extreme view for the R algorithm
would be that after each microstep, for example $\Iota_u \tmop{II}_u$ in
(\ref{Rit}), we in this way measure all $\mathcal{O}(x)$ (mostly implicit zero
contributions). Then the usual division holds and we could obtain
$\tau_{\tmop{int}}$ in units of microsteps, but would need to handle lots of
data for this purpose. In practice we block however O($V$) such measurements
and then both the variance of the blocks and the residual autocorrelations
between them depend on the underlying algorithm's `decorrelation power'.

\begin{figure}[htb]
\begin{center}
  \resizebox{!}{0.8\textwidth}{\includegraphics{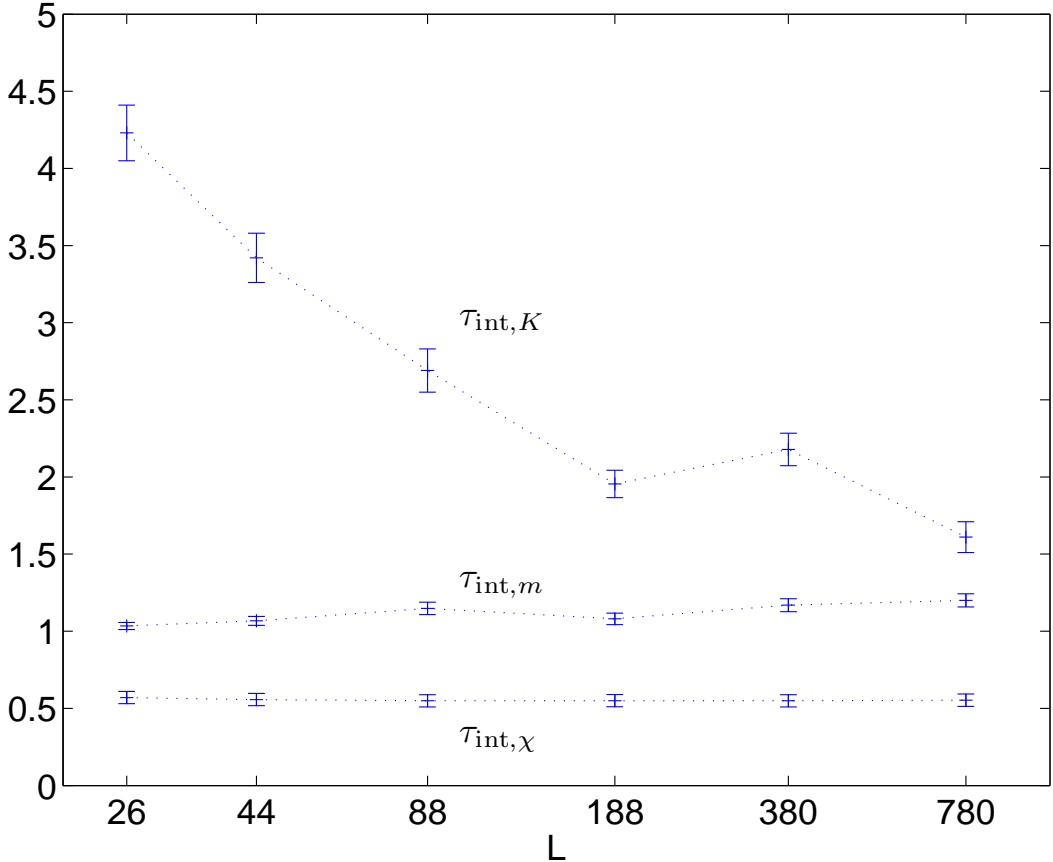}}
  \caption{Integrated autocorrelation times in units of
  I-iterations.\label{fig2}}
\end{center}
\end{figure}

With this explained we cite some autocorrelation times for the I algorithm. To
that end we have repeated the simulations of table \ref{tab2} with only $10^6$
iterations but storing the contribution for each of them separately (in
particular all $G (t)$). Then an ordinary analysis {\cite{Wolff:2003sm}}
yields autocorrelation times in units of iterations (computational complexity
O($V N$), like `sweeps'). Results are given in figure \ref{fig2}. The dotted
lines just connect data points for the same observable. Here
$\tau_{\tmop{int}, m}$ refers to the fitted masses. Values based on
$m_{\tmop{eff}} (L / 4)$ cannot be shown as they would fall on top of
$\tau_{\tmop{int}, \chi}$, i.e. be close to 1/2 for all lattices which means
no autocorrelations in our definition of $\tau_{\tmop{int}}$.
We here see the complete absence of any growth of $\tau_{\tmop{int}}$ as the
continuum limit is taken. The slowest modes couple to the total graph-size $K$
which however even speeds up in the continuum limit. In any case, the most
naive (and important) conclusion from table \ref{tab2} is simply, that we see
an approximately constant relative error in $\xi$ with costs only growing
linearly with the number of sites.

\subsection{Universality and finite size scaling}

We conducted another series of tests for the $N = 3$ model in a finite size
scaling situation. To compare with data in {\cite{Beard:2004jr}} we here
switch to the second moment definition of the mass $m_2$. It is based on
ratios of momentum space correlations which can be measured in the loop
ensemble by
\begin{equation}
  \tilde{G} (p) \propto \langle \langle \rho (u - v) \cos (p \cdot (u - v))
  \rangle \rangle
\end{equation}
which follows from Fourier transforming (\ref{corr2}). Table \ref{tab3}
contains our new data.
\begin{table}[htb]
\begin{center}
  \begin{tabular}{|l|l|l|l|l|}
    \hline
    act. & $\beta_{\times}$ & $m_2 (L) L |_{L = 32}$ & $m_2 (L) L |_{L = 64}$
    & $m_2 (L) L |_{L = 128}$\\
    \hline
    $S_q$ & 2.25 & 2.3979(18) & 3.6582(22) & 7.2640(30)\\
    \hline
    $S_q$ & 2.4 & 1.9896(18) & 2.4133(19) & 3.6631(24)\\
    \hline
    $S_q$ & 2.5 & 1.8351(18) & 2.1080(19) & 2.6905(21)\\
    \hline
    $S_q$ & 3.0 & 1.4285(19) & 1.5286(20) & 1.6575(20)\\
    \hline
    $S$ & 4.0 & 2.1138(17) & 2.6974(19) & 4.6179(26)\\
    \hline
    $S$ & 4.2 & 1.9366(17) & 2.2812(19) & 3.1972(22)\\
    \hline
    $S$ & 4.5 & 1.7499(18) & 1.9646(19) & 2.3433(20)\\
    \hline
    $S$ & 5.0 & 1.5448(18) & 1.6747(19) & 1.8607(20)\\
    \hline
  \end{tabular}
  \caption{Finite volume results for $L = 32, 64, 128$ with actions (\ref{Sq})
  and (\ref{SU}).\label{tab3}}
\end{center}
\end{table}

\begin{figure}[htb]
\begin{center}
  \resizebox{!}{0.8\textwidth}{\includegraphics{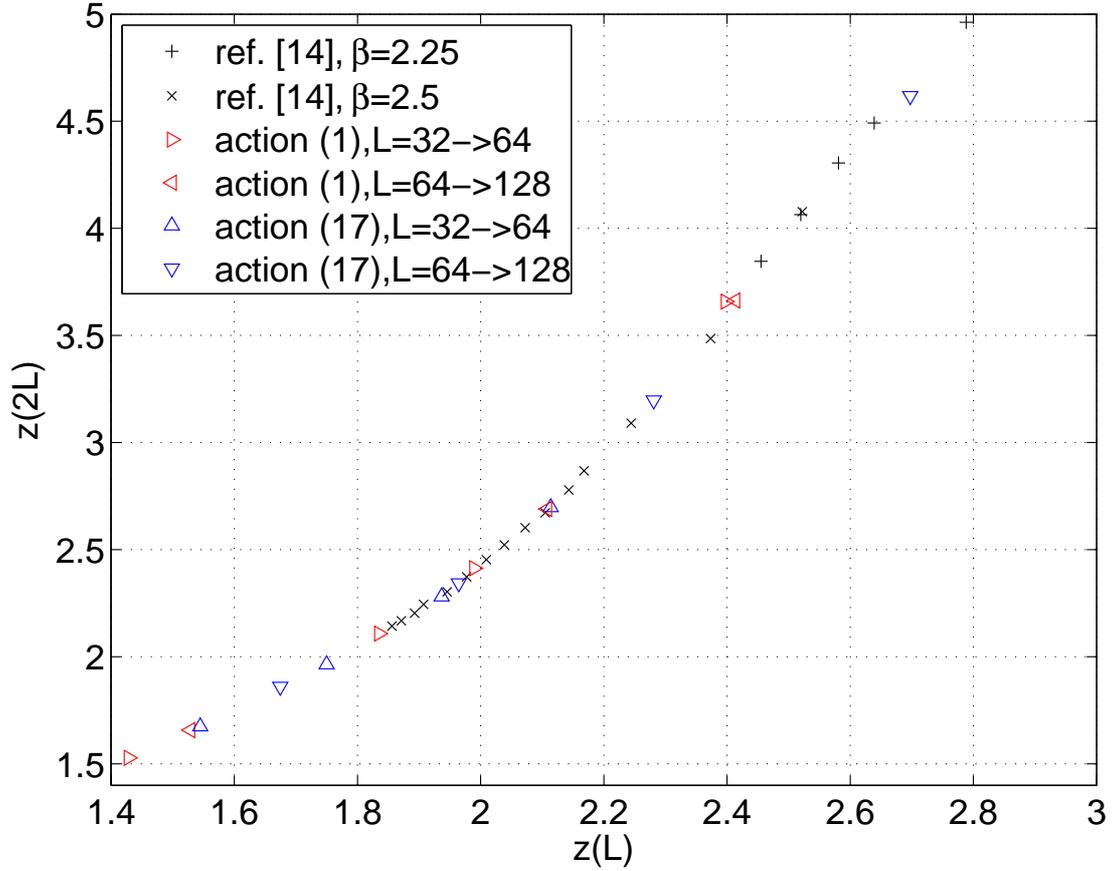}}
  \caption{Finite size scaling function of the CP(2) model with data from two
  different actions and spin as well as loop simulations. Errors (horizontal
  and vertical) are at most comparable to the symbol sizes.\label{fig3}}
\end{center}
\end{figure}

We consider the step scaling function {\cite{Luscher:1991wu}} $z (2 L)$ versus
$z (L)$ for
\begin{equation}
  \text{$z (L) = m_2 (L) \times L$}
\end{equation}
with pairs ($L, 2 L$) at the same $\beta$. For large enough $L$ this graph is
expected to reach a universal continuum curve. Rather then tuning the smaller
lattices as in {\cite{Wolff:2009kp}} we here give only a more qualitative
`curve-collapsing' demonstration. In figure \ref{fig3} we show the pairs
contained in table \ref{tab3} together with data{\footnote{I would like to
thank the authors of {\cite{Beard:2006mq}} for sending their data and allowing
me to reproduce them here.}} from refs. {\cite{Beard:2004jr}},
{\cite{Beard:2006mq}} using their lattice pairs between ($32, 64$) and ($104,
208$) produced with standard simulations using $S_q$ at $\beta_q = 2.25, 2.5$.
We see a very convincing close to universal curve. Some remaining
`roughness' from lattice artefacts expected at a level $L^{- 2}$ can just
still be anticipated, for example around $z (L) \approx 2.4$.

\section{Topological charge in the loop model}

In two dimensions the gauge field $U (x, \mu)$ gives rise to an integer
topological charge $Q$. It has a straight-forward definition on the lattice
and we may hence extend the action by the $\theta$ term by including the phase
$\exp (i \theta Q)$ with the $U$ integrations.

We parameterize
\begin{equation}
  U (x, \mu) = \mathe^{i A_{\mu} (x)}, \hspace{1em} A_{\mu} (x) \in (- \pi,
  \pi]
\end{equation}
and define the field strength
\begin{equation}
  F_{\mu \nu} = \partial_{\mu} A_{\nu} - \partial_{\nu} A_{\mu} \in (- 4 \pi,
  4 \pi] .
\end{equation}
An integer topological charge density $n_{\mu \nu}$ is then extracted by
setting
\begin{equation}
  F_{\mu \nu} = [F_{\mu \nu}] + 2 \pi \text{$n_{\mu \nu}$}, \hspace{1em}
  [F_{\mu \nu}] \in (- \pi, \pi]
\end{equation}
and the global charge is given by
\begin{equation}
  Q = \sum_x n_{01} (x) .
\end{equation}
We next Fourier expand
\begin{equation}
  \mathe^{i \overline{\theta} [F_{01}]} = \sum_{h = - \infty}^{\infty} H (
  \overline{\theta} ; h) \mathe^{i h [F_{01}]} = \sum_{h = - \infty}^{\infty}
  H ( \overline{\theta}, h) \mathe^{i h F_{01}}
\end{equation}
with $\overline{\theta} = \theta / (2 \pi)$ and
\begin{equation}
  H ( \overline{\theta} ; h) = \frac{\sin [( \overline{\theta} - h) \pi]}{(
  \overline{\theta} - h) \pi} = \frac{\sin (\theta / 2)}{\theta / 2} (- 1)^h 
  \frac{\theta}{\theta - 2 \pi h} .
\end{equation}
We remark that $H$ is precisely the kernel that appears in connection with the
sampling theorem {\cite{Press:1058314}}. There a continuous time signal with
compact support in frequency space is reconstructed from its values at
equidistant discrete times. In our case $\overline{\theta}$ is continuous and
the critical Nyquist sampling frequency is one in our units. If
$\overline{\theta}$ gets close to an integer $k$ we have
\begin{equation}
  H (k + \epsilon ; h) = \delta_{k h}
\end{equation}
in a distribution sense [first insert $H$ into a sufficiently convergent sum,
then take $\epsilon \rightarrow 0$]. If we now promote $h$ to a field $h (x)$,
use $\sum_x F_{\mu \nu} = 0$ and rearrange the $U (x, \mu)$ from plaquettes
into in a link-wise order we arrive at
\begin{equation}
  \mathe^{i \theta \sum_x n_{01} (x)} = \sum_h \left[ \prod_x H (
  \overline{\theta} ; h (x)) \right] \prod_{x \mu} [U (x, \mu)]^{-
  \varepsilon_{\mu \nu} \partial_{\nu}^{\ast} h (x)}
\end{equation}
where $\varepsilon_{\mu \nu}$ is the antisymmetric tensor with
$\varepsilon_{01} = + 1$.

The expression
\begin{equation}
  \Theta = \int D U \mathe^{i \theta Q} \prod_{x \mu} U^{j (x, \mu)} = \sum_h
  \left[ \prod_x H ( \overline{\theta} ; h (x)) \right] \prod_{x \mu}
  \delta_{j_{\mu} (x), \varepsilon_{\mu \nu} \partial_{\nu}^{\ast} h (x)}
  \label{Theta}
\end{equation}
is now well-prepared for its use in (\ref{WU}) and leads to
\begin{equation}
  \mathcal{Z}_{\theta} = \sum_{\Lambda \in \overline{\mathcal{L}}_2} \rho^{-
  1} (u - v) W [\Lambda] N^{| \Lambda |} \Theta [\Lambda],
\end{equation}
where $\Theta$ depends on the integer flux $j (x, \mu)$ which we here regard
as a function of $\Lambda$. The set $\overline{\mathcal{L}}_2$ is such that
flux conservation (\ref{divj}) holds, which is a necessary condition for the
constraint in $\Theta$ to have solutions.

In {\cite{Affleck:1991tj}} a very nice discussion is given for a closely
related representation of $\sigma$-models which we adopt now. The field $h
(x)$ is really associated with plaquettes, or sites in the dual lattice. Its
integer values in an $h$ configuration may be viewed as the height of a tower
above (or below) its plaquette, like a `lego-plot' of the experimentalists.
Then the euclidean plane is decomposed into domains of equal height. Due to
the constraint in (\ref{Theta}) the links along their boundaries carry
nonvanishing conserved flux $j (x, \mu)$ whose value and sign is determined by
the difference between the domains heights on both sides. As noted in
{\cite{Affleck:1991tj}} this is a solid-on-solid model with additional
interactions. In the limit $\theta \rightarrow 0$ only $h (x) = 0$ survives.
At $\theta = \pi$ on the other hand there is a reflection symmetry $h (x) - 1
/ 2 \leftrightarrow - (h (x) - 1 / 2)$.

We leave it to a future investigation to decide if or at which ($\beta,
\theta$) in spite of the sign oscillations in the weight $H$ it is possible to
numerically investigate topology in the loop version of the CP($N - 1$) model.

\section{Conclusions}

We have constructed a loop re-formulation of the CP($N - 1$) model which
required just a rather straight-forward extension of the steps used in the
O($N$) model in {\cite{Wolff:2009kp}}. With only minor changes we could obtain
the transcription for both the quartic lattice action and for the formulation
involving an additional U(1) gauge field. For the numerical simulation an
important refinement was necessary to avoid critical slowing down in the CP($N
- 1$) case: the stochastic rather then exact treatment of the weight $N^{|
\Lambda |}$ where $| \Lambda |$ is the number of closed loops in an arbitrary
strong coupling graph. The exact algorithm could still be used to simulate the
theory also at non-integer $N$, if desired, although with accepting some
critical slowing down.

One of the main reasons to invent and study the CP($N - 1$) models was their
analogy with Yang-Mills theory in four dimensions with regard to both
asymptotic freedom and the existence of an integer winding number. As we have
argued that the loop model is an exact representation of the original model
the observables related to topology should possess an image in the loop model.
We have constructed the effect of the term in the action related to the
$\theta$ parameter. It leads however to the appearance of negative weights in
the loop model. It is deferred to future work to find out in which parameter
range, by including the signs in observables or by further re-formulation,
numerical calculations are possible.

{\noindent}\tmtextbf{Acknowledgments}: I would like to thank Tomasz Korzec and
Peter Weisz for discussions and Burkhard Bunk and Stefan Schaefer for advice
with computing and C. Financial support of the DFG via SFB transregio 9 is
acknowledged.

\end{document}